\documentclass[epj,twocolumn]{webofc}
\usepackage[varg]{txfonts}   
\usepackage{url}
\woctitle{LHCP 2013}
\begin{document}
\title{SUSY after LHC8: a brief overview}
%
%

\author{Enrico Bertuzzo\inst{1}\fnsep\thanks{\email{enrico.bertuzzo@cea.fr}} 
}

\institute{Institut de Physique Th\'eorique, CEA-Saclay,
F-91191 Gif-sur-Yvette Cedex, France. 
          }

\abstract{%
  With the 8 TeV LHC run now concluded, the first consequences of the experimental results on the supersymmetric parameter space can be 
  drawn. On one hand, the negative direct searches place more and more stringent bounds on the mass of supersymmetric particles; 
  on the other hand, the discovery of a 125 GeV Higgs boson points toward a quite heavy spectrum for the squarks of the third generation, 
  at least in the minimal supersymmetric model. In this note I will briefly recap how this constitutes a problem for the naturalness of supersymmetric models, 
  as well as 
  the current experimental situation. Moreover, I will point out 
  possible non minimal models in which the naturalness issue can be at least soften.
}
\maketitle
\section{Introduction: the Hierarchy problem, once more}
\label{intro}
  After the discovery of a scalar particle with mass around $125$ GeV and properties closely resembling those of the Standard Model (SM) 
  Higgs boson \cite{Aad:2012tfa, Chatrchyan:2012ufa}, the long standing problem of the 
  stability of fundamental scalar masses has become more acute than ever. 
  As is well known, the problem stems from the Renormalization Group Equation (RGE) for the scalar squared mass parameter. At one loop, it can always 
  schematically be written as
  \begin{equation}
  \label{eq:RGE_ms}
   \frac{d m_S^2}{d\log\mu} = \beta M^2 \, .
  \end{equation}
  Here $M$ is the mass of any particle that enters in the loop correction to the scalar two point function, and $\beta$ a coefficient (calculable 
  in perturbation theory) that encodes informations on the couplings between the scalar and the particles entering in the loop. 
  In the pure SM, the largest contribution is given by the top quark, being it the particle most strongly coupled to the Higgs boson. 
  If we assume Eq.~(\ref{eq:RGE_ms}) to be a fair way to mimic the effect of any new physics appearing above the Electroweak (EW) scale, 
  $M \sim \Lambda_{NP}$ and we expect $m_S^2$ to get corrections proportional to $\Lambda_{NP}^2$:
  \begin{equation}
  \label{eq:delta_mS}
   \delta m_S^2(\Lambda_{EW}) \simeq \beta \Lambda_{UV}^2 \log\left(\frac{\Lambda_{UV}}{\Lambda_{EW}}\right) \, .
  \end{equation}
  That such corrections are present is at least plausible, since we expect the SM to be leastwise somehow coupled to gravity 
  (in which case $\Lambda_{NP} \sim M_{PL}$), but the result is more general and is true for any scale of new physics present above $\Lambda_{EW}$. 
  Assuming the new experimental scalar to be the SM Higgs boson, this result can be problematic, since its mass is observed to be of order of the EW scale and not 
  much larger, as expected from the previous considerations. This is the well known Hierarchy Problem.
  
  Inspection of Eqs.~(\ref{eq:RGE_ms}-\ref{eq:delta_mS}) shows possible ways to evade the problem: we can either assume a very low cut-off $\Lambda_{NP}$, or we can try to 
  cancel all the large contributions adding new particles with suitably chosen couplings.
  The first way has been pursued in the last years in models of Composite Higgs (see~\cite{Contino:2010rs} for a review), in which moreover 
  the Higgs boson is required to be a Pseudo-Goldstone boson 
  of some dynamically broken group in order to guarantee its lightness. The second way is pursued, at least up to a certain extent, 
  in Supersymmetric (SUSY) theories, on which we are going to focus in the following.
  
  The remaining of this note is organized as follows: In Sec.~\ref{sec:SUSY} SUSY will be briefly discussed, especially in connection with the 
  LHC phenomenology and the fine tuning problem. In Sec.~\ref{sec:LHC8} the latest LHC results will be discussed, together with their consequences on Supersymmetric 
  parameter space. In Sec.~\ref{sec:alternatives}, insisting on naturalness, some alternatives to the standard SUSY scenario are presented (with no attempt 
  to completeness, given the huge number of relevant alternatives presented in the literature). Some conclusions and perspectives are presented in 
  Sec.~\ref{sec:conclusions}.
  
\section{Supersymmetry and Naturalness}
\label{sec:SUSY}

The structure of supersymmetric theories is too well known to be discussed in detail here. \footnote{For a comprehensive review, 
see~\cite{Martin:1997ns}} Let us just stress the features essential for our discussion. As far as exact SUSY is assumed, in Eq.~(\ref{eq:RGE_ms}) only 
a wave function renormalization contribution proportional to $m_S^2$ itself survives. In this case, $m_S^2$ is protected from large radiative correction. 
However, it is well known that in any SUSY theory which attempts to be phenomenologically relevant, SUSY must be broken. 
The presence of soft SUSY breaking terms in the lagrangian 
modifies once more the RGE for any scalar squared mass, in particular those relevant for the Higgs boson mass. Schematically, they can now be written as
\begin{equation}
\label{eq:RGE_SUSY}
 \frac{d m_h^2}{d\log\mu} = \beta m_{SUSY}^2 \, ,
\end{equation}
where $m_{SUSY}^2$ is the scale of SUSY breaking. Apparently, we are back to the original problem, Eq.~(\ref{eq:RGE_ms}). However, the hope is now that 
the scale $m_{SUSY}^2$ of Eq.~(\ref{eq:RGE_SUSY}) is low enough for the theory to be only mildly tuned.
A useful way to measure the amount of fine tuning is given by~\cite{Barbieri:1987fn, Dimopoulos:1995mi}
\begin{equation}
 \Delta = \frac{\partial \log m_h^2}{\partial \log a^2} \, .
\end{equation}
Here, $a^2$ is any soft parameter that enters in the determination of the Higgs mass (or equivalently, of the EW scale), and $1/\Delta$ measures the amount of 
tuning of the theory. For instance, $\Delta = 100$ implies a tuning at the percent level.

We can now draw some conclusions on the typical mass scales expected for SUSY particles. Using common notation, we obtain~\cite{Papucci:2011wy}:
\begin{eqnarray}
\label{eq:natural_spectrum}
 \mu &\lesssim& 200\;{\rm GeV}\left(\frac{m_{h}}{125\;{\rm GeV}}\right)\sqrt{\frac{\Delta}{5}} \; ,\nonumber\\
\overline{m}_{\tilde{t}}&\lesssim & 600\;{\rm GeV} \frac{\sin\beta}{\sqrt{1+x_{t}^{2}}}
{\cal F}(\Lambda) \left(\frac{m_{h}}{125\;{\rm GeV}}\right)\sqrt{\frac{\Delta}{5}} \; , \nonumber \\
M_{3} &\lesssim & 900\;{\rm GeV} \sin\beta \; {\cal F}^2(\Lambda) \left(\frac{m_{h}}{125\;{\rm GeV}}\right)\sqrt{\frac{\Delta}{5}} \; , \nonumber\\
\end{eqnarray}
where $\overline{m}_{\tilde{t}}=\sqrt{ m_{\tilde{t}_1}^2+ m_{\tilde{t}_2}^2}$ is the averaged stop mass, $x_t=A_t/\overline{m}_{\tilde{t}}$ measures the 
left-right mixing in the stop sector and ${\cal F}(\Lambda) = \sqrt{\frac{3}{\log\left(\Lambda/{\rm TeV}\right)}}$ gives the 
logarithmic factor coming from the leading-log solution of the RGE, normalized to a scale $\Lambda\simeq 20$ TeV. 
All the other s-particles can be above the TeV scale without introducing too 
much fine tuning~\cite{Barbieri:2010pd}, with the exception of the left handed scalar bottom which, 
being related to the stops by gauge symmetry, shares the same bound.

As a general message, we see that squarks of the third generation (the one more strongly coupled to the Higgs system) has to be below 
the TeV scale to have a natural theory, preferably with small left-right mixing. At the same time, the gluino mass $M_3$ can be around the TeV scale, but not much heavier. 
The Higgsino mass $\mu$, on the other hand, must be in the $200$ GeV ballpark: such a low mass scale is due to the fact that this parameter enters in the 
determination of the Higgs mass already at tree level, and is thus 
more constrained than the parameters entering only at loop level.

\section{SUSY facing LHC8}
\label{sec:LHC8}

\begin{figure*}
\centering
\includegraphics[width=.9\textwidth,clip]{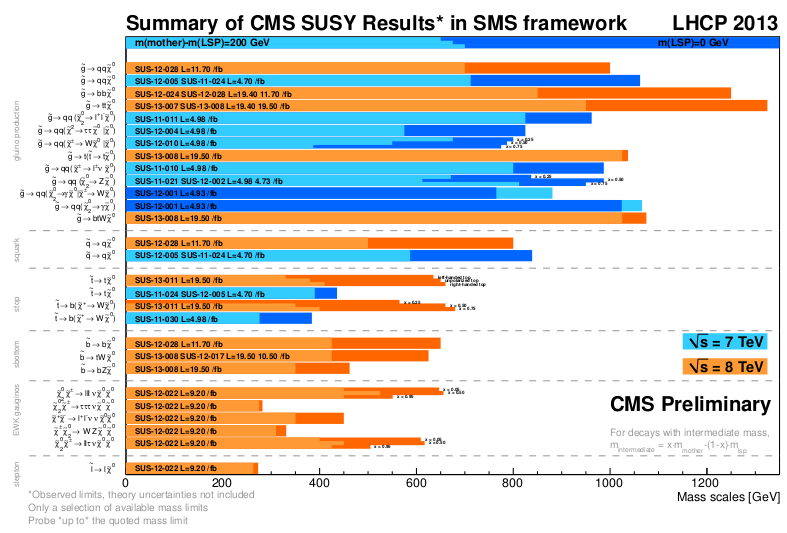}
\caption{Updated limits on supersymmetric particles masses obtained by the CMS collaboration. Similar bounds are obtained by the Atlas collaboration. 
See \cite{simplified_exclusion_Atlas, simplified_exclusion_CMS} for details.}
\label{fig:CMS_limits}       
\end{figure*}
Let us now comment on the experimental results which, as is well known, have given negative results on the direct search for supersymmetric particles. 
In the past years, experimental results on SUSY parameter space were given in the CMSSM/mSUGRA framework, in which all the three generations of squarks are basically 
degenerate at the weak scale. The advantage of this scenario (far from being the most general one) is the limited number of independent parameters defined 
at high energy: $m_0$, 
$m_{1/2}$, $A_0$, $\tan\beta$ and ${\rm sign}(\mu)$. Exclusion limits are usually casted fixing the last three parameters. For instance, 
fixing $A_0=-2m_0$, $\tan\beta=30$ and $\mu>0$ the limits on gluino and squark masses are \cite{ATLAS-CONF-2013-062, ATLAS-CONF-2013-054}
\begin{equation}
 m_{\tilde{g}} \gtrsim 1.4\;{\rm TeV}, ~~~~~~m_{\tilde{q}} \gtrsim 1.7\;{\rm TeV}
\end{equation}
These limits are mainly driven by the strong production of squarks of the first two generations. In the general case, we expect limits on squarks of the 
third generation to be weaker, given the smaller production cross section at the LHC. 

The way experiments choose to present limits for a generic 
spectrum is in terms of 
``simplified models'' in which, for given production and decay modes, only the relevant degrees of freedom are retained in the spectrum \cite{Chatrchyan:2013sza}. 
An updated set of limits can be found in the Atlas and CMS websites \cite{simplified_exclusion_Atlas, simplified_exclusion_CMS}. The general 
message is that while the limit on the gluino mass is roughly comparable with the one obtained in the constrained-MSSM scenario, limits on stops and sbottoms 
are weaker (see Fig.~\ref{fig:CMS_limits}):
\begin{equation}
 m_{\tilde{g}} \gtrsim 1.2\;{\rm TeV}, ~~~m_{\tilde{t}, \tilde{b}} \gtrsim (600-700)\;{\rm GeV} \; .
\end{equation}
As can be seen, direct searches are exploring the region relevant for naturalness.

On top of this, also the results on the Higgs mass are relevant for our considerations. It is well known that in minimal SUSY models the tree level 
Higgs boson mass is bounded from above by the Z mass, so that large radiative correction are needed to increase the physical mass up to the observed value. 
The most relevant corrections are given by stops, \emph{i.e.} exactly the same particles relevant for naturalness. In particular, it turns out that stops 
in the multi-TeV range and/or large stop mixing is needed to achieve a $125$ GeV Higgs boson. 
This has to be compared with Eq.~(\ref{eq:natural_spectrum}): the immediate consequence is that we expect \cite{Hall:2011aa}
\begin{equation}
 \Delta \gtrsim 100\; ,
\end{equation}
\emph{i.e.} the minimal models are tuned \emph{at best} at the percent level.

There is no firm theorem stating that this amount of tuning is unacceptable. However, if naturalness is chosen at all by Nature as a guiding principle, 
we may expect the level of fine tuning to be less pronounced, possibly in the $\Delta\simeq 10-20$ ballpark. We see then that, despite the negative 
results from direct searches, the most problematic result for naturalness is the measured Higgs boson mass.

Different ways of thinking have been put forward in the literature concerning the fine tuning problem in the SM and in minimal SUSY. They can be broadly 
summarized as follows:
\begin{itemize}
 \item The way of fine tuning (strong version): we accept the SM as it stands. The lightness of the Higgs boson is due to tuning among parameters but no new 
 physics is present (apart from an explanation of neutrino masses and Dark Matter);
 \item The way of fine tuning (milder version): we accept that most of the Hierarchy problem is cured by SUSY, but the TeV-scale soft SUSY breaking terms 
 cause the theory to be tuned at best at the percent level;
 \item The way of naturalness: we take naturalness as a principle followed by Nature, searching for minimally tuned theories (we can for instance set a limit 
 of possible acceptable fine tuning at the level of $5-10\%$, although this is largely arbitrary). 
 Sticking to SUSY, this motivates the search for models beyond the minimal one.
\end{itemize}
Clearly, taking the first two point of views, we do no need to do anything beyond the minimal theories apart from possibly refining calculations in such 
frameworks (of course this is a bit reductive, since for instance we lack of any kind of explanation concerning the SUSY flavor problem or, more in general, 
there is no established theory of flavor). On the other hand, 
some model building is necessary taking the third point of view, as explained in Sec.~\ref{sec:alternatives}.

\section{Alternatives to the ``standard'' scenario}
\label{sec:alternatives}

We take in this section the following point of view: naturalness is a good guiding principle of Nature, and we thus seek for minimally tuned theories. 
Taking SUSY as starting point, this motivates the search for SUSY theories beyond the minimal one. 
Without any attempt to completeness, we can try to classify possible proposals as follows:
\begin{enumerate}
 \item ``Hiding stops'': the stops are in the natural region, but they are not detected because hidden in kinematically difficult regions;
 \item ``Increasing $m_h^2$ at tree level'': if the Higgs boson mass is increased at tree level, the sensitivity of its mass to radiative corrections is 
 diminished, improving naturalness;
 \item ``Mixing stops'': a mild improvement in the amount of fine tuning is achieved mixing the stops with other squarks. In particular, mixing with the scalar charm 
 allows the flavor bounds, as well as direct searches, to be satisfied;
 \item ``Adding stops'': additional ``stops'' (\emph{i.e.} particles that give a radiative boost to the Higgs quartic comparable to the stops one) are added in the 
 theory. Possibly, multiple loop contributions (each one natural) can increase the Higgs boson mass up to the observed level without introducing too much 
 fine tuning in the theory.
\end{enumerate}

\subsection{Hiding Stops}
\label{subsec:hiding_stops}

Stops and sbottoms searches in simplified models typically consider the decay channels $\tilde{t} \rightarrow t \chi^0$ or $\tilde{b} \rightarrow b \chi^0$, with the 
sensitivity diminishing when the spectrum becomes compressed, \emph{i.e.} when $m_{\tilde{t}, \tilde{b}} \sim m_{t,b} + m_{\chi^0}$ (see Fig.~\ref{fig:CMS_limits}). 
\footnote{See also~\cite{Dreiner:2012gx} and references therein.}
A possibility is thus that stops and sbottoms are in the natural region but are not detected because the spectrum is somewhat compressed. 
If this is so, on rather general ground we can expect the next run of the LHC to shed complete light on this region. In any case, if both stops 
are hidden in this region, in the minimal model there is no way to increase the Higgs boson mass up to the observed level, so that 
additional structures have anyway to be added (see Sec.~\ref{subsec:increasing_mh}). The 
other possibility is to have one light and one heavy stop, with the light one hidden in the compressed region and the other one responsible 
for the Higgs boson mass. However, taking naturalness as guiding principle, this is not a viable solution since a large stop mixing is needed, increasing 
in this way the fine tuning of the theory.

\subsection{Increasing $m_h^2$ at tree level}
\label{subsec:increasing_mh}

As already pointed out, the supersymmetric naturalness problem is a direct consequence of the tree level upper bound on the Higgs boson mass, 
with sizeable radiative corrections needed to increase its value up to 125 GeV. It is clear that the situation 
can be ameliorated introducing in the theory additional structures to increase $m_h$ already at tree level. There are three possibilities:
\begin{itemize}
 \item F-terms: the Higgs quartic coupling is increased acting on the F-term part of the tree level scalar potential. 
 This is achieved introducing new supersymmetric couplings between the two Higgs doublets and additional superfields in the superpotential. The new 
 superfield can either be an electroweak singlet (as in the NMSSM/$\lambda$SUSY, see~\cite{Ellwanger:2009dp, Maniatis:2009re, Barbieri:2006bg} and 
 references therein), or one or more triplet (see~\cite{DiChiara:2008rg, FileviezPerez:2012ab} and references therein);
 \item D-term: the Higgs quartic is increased adding a D-term to the tree level scalar potential, \emph{i.e.} extending the SM gauge group. Possible 
 examples are the gauge groups SU(3)$\times$SU(2)$\times$U(1)$\times$U(1) or SU(3)$\times$SU(2)$\times$SU(2)$\times$U(1)~\cite{Batra:2003nj, Barbieri:2010pd};
 \item Non renormalizable terms: if the scale of SUSY breaking is very low, in the few-TeV range, then non renormalizable operators in the 
 Kahler potential can give relevant contribution to the Higgs quartic, see for example~\cite{Brignole:2003cm, Casas:2003jx}.
\end{itemize}
Once the tree level Higgs boson mass is increased, it is clear that the sensitivity of the physical mass to radiative correction is 
diminished, allowing in particular for a heavier 
stops spectrum without introducing too much fine tuning.

\subsection{Mixing Stops}
\label{subsec:mixing_stops}

A mild improvement in the fine tuning of the MSSM can be achieved assuming a large mixing between the right-handed stop and the right-handed 
scharm~\cite{Blanke:2013uia, Agrawal:2013kha}. The main observation stems from the fact that, in presence of such a mixing, the RGE for the Higgs 
mass parameter are modified by the $\tilde{t}_R-\tilde{c}_R$ mixing. \footnote{The usual RGE for the Higgs mass parameter in absence of 
mixing reads $\frac{dm_h^2}{dt} = -\frac{3 y_t^2}{8\pi^2}\left(m^2_{\tilde{t}_L} + m^2_{\tilde{t}_R} + |A_t|^2\right)$. When the $\tilde{t}_R-\tilde{c}_R$ mixing 
is included, the previous equation must be modified by $m^2_{\tilde{t}_R} \rightarrow c^2 m_1^2 + s^2 m_2^2$. 
Here $c=\cos\theta^{tc}$, $s=\sin\theta^{tc}$ (with $\theta^{tc}$ the stop-scharm mixing angle) and $m_{1,2}$ are respectively the 
masses of the mostly stop and mostly scharm states.} This scenario is more favorable than the usual one since the s-charm mass is still only mildly constrained 
by LHC direct searches. At the same time, a mixing in the right handed sector is safe from flavor constraints. The neat effect is to ameliorate the fine tuning due to 
the right handed part of the spectrum by roughly 30\%.

\subsection{Adding stops}
\label{subsec:adding_stops}

A further possibility in trying to ameliorating the supersymmetric fine tuning issue is to assume the existence of a non minimal particle content, with 
 additional states giving a sizeable radiative correction to the Higgs boson mass (with contributions roughly of the same order 
as the stops ones). The difference with respect to the situation outlined in Sec.~\ref{subsec:increasing_mh} is that here there is no tree level enhancement 
of the Higgs boson mass. This kind of situation naturally arises in models with an approximate R-symmetry~\cite{Hall:1990hq, Fox:2002bu, Kribs:2007ac, Davies:2011mp, 
Frugiuele:2011mh, Frugiuele:2012pe, Frugiuele:2012kp}, where the role of additional ``stops'' may be played by the adjoint superfields added to 
give Dirac masses to the gauginos. 
A possibility is thus to increase the Higgs boson mass up to the experimental value using multiple loop contributions coming from all the 
``stop'' particles present in the theory (the stops themselves and the gaugino partners), with soft masses for each particle kept 
small enough in order not to introduce too much fine tuning in the theory~\cite{adding_stops}.

\section{Conclusions}
\label{sec:conclusions}

With the first LHC run now concluded, we can start drawing conclusions on the shape of the theory of particle physics. 
The main result that emerges from colliders (and from flavor physics) is the impressive agreement between the Standard Model and experimental data. 
If any new physics is present, it is likely 
to manifest itself only as a small perturbation around the Standard Model expectations. While from the experimental point of view this is perfectly fine 
(and actually a great achievement), from the theoretical point of view many points still remain open. Particularly pressing is the question 
of how the Higgs boson mass is kept light, since radiative corrections are likely to drive it to much larger values than the one observed. This basic observation 
has motivated much of the theoretical activity in the last decades, leading to many proposals of completions of the Standard Model 
designed to give a naturally light electroweak scale. 
The case of supersymmetry is particularly emblematic of the current situation of most of such theories: although designed to give 
a naturally light Higgs boson in a large part of their parameter space, current experimental data are probing more and more the natural region, with negative results. 
As a consequence, a partial conclusion can already be drawn: either Nature is simple but fine tuned, or it is natural but not simple. 
While the search for natural theories is currently still motivated (some examples were presented in Sec.~\ref{sec:alternatives} in a supersymmetric contest), 
it may be that if no sign of new physics will emerge in the next run of the LHC (or will emerge well outside the natural region), 
the concept of naturalness may have to be definitively abandoned.

\section*{Acknowledgments}
It is a pleasure to thank the organizers of LHCP 2013. I am indebted to Riccardo Barbieri, Marco Farina, Claudia Frugiuele, 
Thomas Gr\'egoire, St\'ephane Lavignac, Paolo Lodone, Eduardo Pont\'on and Carlos A. Savoy for many stimulating discussions on the 
topics here presented. This work was supported by the Agence National de la Recherche under contract ANR 2010 BLANC 0413 01.

\bibliography{SUSY_LHC8}

\begin{thebibliography}{33}

\bibitem{Aad:2012tfa}
G.~Aad et~al. (ATLAS Collaboration), Phys.Lett. \textbf{B716}, 1 (2012),
  \texttt{1207.7214}

\bibitem{Chatrchyan:2012ufa}
S.~Chatrchyan et~al. (CMS Collaboration), Phys.Lett. \textbf{B716}, 30 (2012),
  \texttt{1207.7235}

\bibitem{Contino:2010rs}
R.~Contino (2010), \texttt{1005.4269}

\bibitem{Martin:1997ns}
S.P. Martin (1997), \texttt{hep-ph/9709356}

\bibitem{Barbieri:1987fn}
R.~Barbieri, G.~Giudice, Nucl.Phys. \textbf{B306}, 63 (1988)

\bibitem{Dimopoulos:1995mi}
S.~Dimopoulos, G.~Giudice, Phys.Lett. \textbf{B357}, 573 (1995),
  \texttt{hep-ph/9507282}

\bibitem{Papucci:2011wy}
M.~Papucci, J.T. Ruderman, A.~Weiler, JHEP \textbf{1209}, 035 (2012),
  \texttt{1110.6926}

\bibitem{Barbieri:2010pd}
R.~Barbieri, E.~Bertuzzo, M.~Farina, P.~Lodone, D.~Pappadopulo, JHEP
  \textbf{1008}, 024 (2010), \texttt{1004.2256}

\bibitem{simplified_exclusion_Atlas}
\texttt{{https://twiki.cern.ch/twiki/bin/
  view/AtlasPublic/SupersymmetryPublicResults}}

\bibitem{simplified_exclusion_CMS}
\texttt{{https://twiki.cern.ch/twiki/bin/
  view/CMSPublic/SUSYSMSSummaryPlots8TeV}}

\bibitem{ATLAS-CONF-2013-062}
Tech. Rep. ATLAS-CONF-2013-062, CERN, Geneva (2013)

\bibitem{ATLAS-CONF-2013-054}
Tech. Rep. ATLAS-CONF-2013-054, CERN, Geneva (2013)

\bibitem{Chatrchyan:2013sza}
S.~Chatrchyan et~al. (CMS Collaboration) (2013), \texttt{1301.2175}

\bibitem{Hall:2011aa}
L.J. Hall, D.~Pinner, J.T. Ruderman, JHEP \textbf{1204}, 131 (2012),
  \texttt{1112.2703}

\bibitem{Dreiner:2012gx}
H.K. Dreiner, M.~Kramer, J.~Tattersall, Europhys.Lett. \textbf{99}, 61001
  (2012), \texttt{1207.1613}

\bibitem{Ellwanger:2009dp}
U.~Ellwanger, C.~Hugonie, A.M. Teixeira, Phys.Rept. \textbf{496}, 1 (2010),
  \texttt{0910.1785}

\bibitem{Maniatis:2009re}
M.~Maniatis, Int.J.Mod.Phys. \textbf{A25}, 3505 (2010), \texttt{0906.0777}

\bibitem{Barbieri:2006bg}
R.~Barbieri, L.J. Hall, Y.~Nomura, V.S. Rychkov, Phys.Rev. \textbf{D75}, 035007
  (2007), \texttt{hep-ph/0607332}

\bibitem{DiChiara:2008rg}
S.~Di~Chiara, K.~Hsieh, Phys.Rev. \textbf{D78}, 055016 (2008),
  \texttt{0805.2623}

\bibitem{FileviezPerez:2012ab}
P.~Fileviez~Perez, S.~Spinner, Phys.Rev. \textbf{D87}, 031702 (2013),
  \texttt{1211.1025}

\bibitem{Batra:2003nj}
P.~Batra, A.~Delgado, D.E. Kaplan, T.M. Tait, JHEP \textbf{0402}, 043 (2004),
  \texttt{hep-ph/0309149}

\bibitem{Brignole:2003cm}
A.~Brignole, J.~Casas, J.~Espinosa, I.~Navarro, Nucl.Phys. \textbf{B666}, 105
  (2003), \texttt{hep-ph/0301121}

\bibitem{Casas:2003jx}
J.~Casas, J.~Espinosa, I.~Hidalgo, JHEP \textbf{0401}, 008 (2004),
  \texttt{hep-ph/0310137}

\bibitem{Blanke:2013uia}
M.~Blanke, G.F. Giudice, P.~Paradisi, G.~Perez, J.~Zupan, JHEP \textbf{1306},
  022 (2013), \texttt{1302.7232}

\bibitem{Agrawal:2013kha}
P.~Agrawal, C.~Frugiuele (2013), \texttt{1304.3068}

\bibitem{Hall:1990hq}
L.~Hall, L.~Randall, Nucl.Phys. \textbf{B352}, 289 (1991)

\bibitem{Fox:2002bu}
P.J. Fox, A.E. Nelson, N.~Weiner, JHEP \textbf{0208}, 035 (2002),
  \texttt{hep-ph/0206096}

\bibitem{Kribs:2007ac}
G.D. Kribs, E.~Poppitz, N.~Weiner, Phys.Rev. \textbf{D78}, 055010 (2008),
  \texttt{0712.2039}

\bibitem{Davies:2011mp}
R.~Davies, J.~March-Russell, M.~McCullough, JHEP \textbf{1104}, 108 (2011),
  \texttt{1103.1647}

\bibitem{Frugiuele:2011mh}
C.~Frugiuele, T.~Gregoire, Phys.Rev. \textbf{D85}, 015016 (2012),
  \texttt{1107.4634}

\bibitem{Frugiuele:2012pe}
C.~Frugiuele, T.~Gregoire, P.~Kumar, E.~Ponton, JHEP \textbf{1303}, 156 (2013),
  \texttt{1210.0541}

\bibitem{Frugiuele:2012kp}
C.~Frugiuele, T.~Gregoire, P.~Kumar, E.~Ponton, JHEP \textbf{1305}, 012 (2013),
  \texttt{1210.5257}

\bibitem{adding_stops}
E.~Bertuzzo, C.~Frugiuele, T.~Gregoire, E.~Ponton (2013), \texttt{To appear}

\end{thebibliography}

\end{document}